\documentclass[aps,prb,preprint,groupedaddress,showpacs]{revtex4-1}
\usepackage{graphicx}
\usepackage{amsmath}
\usepackage{amssymb}
\usepackage{dcolumn}
\usepackage{hyperref}

\begin{document}

\title{Influence of tetragonal platelets on the dielectric permittivity of \texorpdfstring{$0.964\textrm{Na}_{1/2}\textrm{Bi}_{1/2}\textrm{TiO}_3-0.036\textrm{BaTiO}_3$}{BNT-3.6BT}}

\author{Florian Pforr}
\author{M{\'a}rton Major}\altaffiliation{On leave from Wigner Research Centre for Physics, RMKI, Budapest, Hungary}
\author{Wolfgang Donner}
\email[]{wdonner@tu-darmstadt.de}
\affiliation{Institute of Materials Science, Technische Universit\"at Darmstadt, Alarich-Weiss-Stra\ss e 2, 64287 Darmstadt, Germany}
\author{Uwe Stuhr}
\author{Bertrand Roessli}
\affiliation{Laboratory for Neutron Scattering and Imaging, Paul Scherrer Institut, 5232 Villigen PSI, Switzerland}

\date{\today}

\begin{abstract}
We study the temperature-dependent evolution of the octahedral tilt order in a lead-free relaxor ferroelectric and its impact on the ferroelectric properties. Using diffuse neutron scattering on a $0.964\textrm{Na}_{1/2}\textrm{Bi}_{1/2}\textrm{TiO}_3-0.036\textrm{BaTiO}_3$ single crystal, we suggest a model for the temperature-dependent nanostructure of this perovskite that features chemically pinned tetragonal platelets embedded in the rhombohedral matrix, often separated by a cubic intermediate phase. Our results show a clear correlation between the squared thickness of the tetragonal platelets and the dielectric permittivity. This is interpreted as a sign for increased polarizability of the strained and distorted lattice at the center of the tetragonal platelets.
\end{abstract}

\maketitle

\section{Introduction}

The scientific quest to replace commonly used lead-based ferroelectrics with lead-free alternatives, which would be significantly less hazardous for human health and our environment, has been ongoing for a number of years.\cite{Roedel2009} The very first customer products using lead-free ferroelectrics are currently being introduced to the market.\cite{Roedel2015} However, the properties of those lead-free materials that are known today need to be improved further in order to make them suitable for a wider range of applications. The solid solution of $\textrm{Na}_{1/2}\textrm{Bi}_{1/2}\textrm{TiO}_3$ (NBT) and $\textrm{BaTiO}_3$ is known as a promising candidate, but the mechanisms leading to its relaxor properties are not yet clear.\cite{Roedel2015} It is thus evident that more research needs to be undertaken in order to clarify the underlying mechanisms and systematically investigate ways to increase the performance of this material.\\

As a solid solution between two perovskite titanates, $(1-x)\textrm{Na}_{1/2}\textrm{Bi}_{1/2}\textrm{TiO}_3-x\textrm{BaTiO}_3$ (NBT-BT) also possesses a perovskite structure.\cite{Jo2011a, Picht2010} Depending on composition and temperature, cubic, tetragonal, rhombohedral, and monoclinic symmetries have been reported.\cite{Takenaka1991, Ma2011, Ma2013} In the specific case $x = 3.6\%$ (NBT-3.6BT), the cubic $Pm\bar{3}m$, tetragonal $P4bm$ and rhombohedral $R3c$ structures have traditionally been assigned as a function of decreasing temperature.\cite{Ma2010, Daniels2011, *Daniels2012} However, it is still unclear whether the original phase diagrams by \textcite{Takenaka1991} on poled ceramics and by Ma and Tan\cite{Ma2010} for unpoled ceramics have to be revised to include a monoclinic $Cc$ phase field.\cite{Gorfman2010, *Gorfman2012, Aksel2011a, Usher2012, Ma2013, Beanland2014, Fuentes-Cobas2014, Ge2014, Gorfman2015} Since a reliable distinction between the rhombohedral and monoclinic phases is not required for our analysis, and because it is experimentally challenging, we will use the designation ``rhombohedral'' instead of ``rhombohedral or monoclinic''.\\

\begin{figure}
\includegraphics[width=6.6cm]{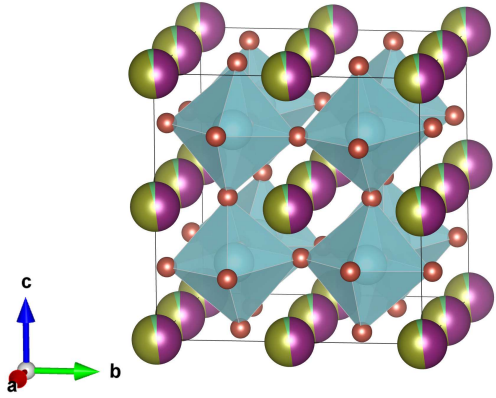}%
\caption{Model of the NBT-3.6BT room temperature structure featuring $a^-a^-a^-$ tilts ($R3c$ symmetry). The $\textrm{Ti}^{4+}$ ions are located at the centers of the oxygen octahedra (B-sites). The A-sites are shared between the $\textrm{Na}^+$, $\textrm{Bi}^{3+}$, and $\textrm{Ba}^{2+}$ cations. The depicted positions of the $\textrm{Bi}^{3+}$ cations are the space and time averages of their true disordered positions. See the text for details. The drawings of the atomic structure were generated using the program VESTA.\cite{Momma2011}\label{fig:atomic_structure}}
\end{figure}

In the space groups $P4bm$, $R3c$, and $Cc$, the pseudocubic unit cell can be distorted in different ways. All of these space groups possess a characteristic polar axis, along which ferroelectric shifts of the cations are possible. Furthermore, these space groups permit tilting of the oxygen octahedra about the respective polar axis. These two order parameters, i.e., ferroelectric shifts and octahedral tilting, can indeed be coupled.\cite{Groeting2013} In any case, the determination of the local tilt system is an established method for the identification of the local symmetry,\cite{Glazer1972, *Glazer1975, Woodward2005} and thus possible domain configurations. In Glazer's notation,\cite{Glazer1972, *Glazer1975} which is commonly used for specifying octahedral tilt systems in perovskites, the tilt system in the cubic phase can be described as $a^0a^0a^0$ (no tilt). The tetragonal phase possesses $a^0a^0c^+$ in-phase tilting about the $c$ axis. Finally, the rhombohedral phase is associated with $a^-a^-a^-$ antiphase tilt. The double pseudocubic unit cell of NBT-3.6BT with $R3c$ symmetry is depicted in Fig.\ \ref{fig:atomic_structure}. The edge length is $2a = 7.8134~\textrm{\AA}$ and the rhombohedral angle is $\alpha = 89.88^\circ$.\cite{Jo2011a} The local disorder of the $\textrm{Bi}^{3+}$ positions\cite{Shuvaeva2005, Thomas2005} is not shown. The $a^0a^0c^+$ tilt system with tetragonal symmetry leads to superlattice reflections (SRs) at $\frac{1}{2}\{ooe\}$-type positions,\cite{Woodward2005} where $o$ stands for odd and $e$ for even integers. From here on, all reflections will be indexed with respect to the single pseudocubic unit cell with $a = 3.9067~\textrm{\AA}$. The $\frac{1}{2}\{ooe\}$-type reflections are associated with the tetragonal phase and will be collectively referred to as ``$T$-type''. Since $\frac{1}{2}\{ooo\}$-type (except for the $R3c$ forbidden $\frac{1}{2}\{hhh\}$, $h$ odd) SRs are due to $a^-a^-a^-$ tilts, they will be designated as ``$R$-type''. The designation ``$M$-type'' will be used for the $\frac{1}{2}\{hhh\}$ ($h$ odd) SRs, which could be caused by a monoclinic $Cc$ structure\cite{Gorfman2010} with $a^-a^-c^-$ tilting.\cite{Stokes2002} The occurrence of mixed tilting, most importantly in the orthorhombic $Pnma$ phase with the $a^-a^-c^+$ tilt system,\cite{Dorcet2008a} would lead to the observation of additional $\frac{1}{2}\{oee\}$-type (except for the $Pnma$ forbidden $\frac{1}{2}\{h00\}$, $h$ odd) SRs,\cite{Woodward2005} designated as ``O-type''.\\

The sequence of phase transitions in NBT-3.6BT is comparable to that of pure NBT. Detailed studies of the phase transitions in the NBT system have been performed during the past 30 years.\cite{Pronin1980, Zvirgzds1982, Vakhrushev1982, Jones2002, Balagurov2006, Dorcet2008a, Trolliard2008, Aksel2011, Deng2014, Gorfman2015} The addition of 3.6\% $\textrm{BaTiO}_3$ shifts the temperature regions of the diffuse phase transitions from 503~K to approximately 463~K ($R3c \rightarrow P4bm$) and from 613 to 583~K ($P4bm \rightarrow Pm\bar{3}m$).\cite{Takenaka1991} Moreover, diffuse scattering studies on both systems have documented the presence of $\frac{1}{2}\{ooe\}$-type SRs close to room temperature.\cite{Kusz1999, Trolliard2008, Daniels2011, *Daniels2012, Yao2012, Matsuura2013, Ge2013} On the other hand, no $T$-type SRs have been observed in recent neutron and synchrotron x-ray powder-diffraction experiments.\cite{Usher2012, Rao2013, *Rao2013a, Garg2013, *Garg2013a, Ge2014, Fuentes-Cobas2014} This can be understood considering that the $\frac{1}{2}\{ooe\}$-type SRs are elongated close to room temperature and exhibit a rather diffuse character,\cite{Trolliard2008, Daniels2011, *Daniels2012, Ge2013} so that the corresponding powder reflections become much broader and less intense than those that would be expected from a bulk tetragonal sample. Additionally, the low x-ray scattering power of oxygen makes the $T$-type SRs particularly weak in x-ray powder diffraction. Consequently, they may not be detectable if the tetragonal phase fraction is too low. Furthermore, \textcite{Ge2012} have shown that $a^0a^0c^+$ octahedral tilt order can occur even when no measurable tetragonal distortion of the crystal lattice is present. In other words, the absence of the splitting of fundamental reflections such as $(200)$ in a powder diffraction pattern does not imply the absence of $a^0a^0c^+$ tilting. Indeed, a cubic minority phase has been proposed by \textcite{Aksel2011} and \textcite{Usher2012} based on their powder diffraction profile refinements. It is thus conceivable that the macroscopically rhombohedral NBT and NBT-3.6BT contain a low volume fraction of a metrically cubic phase with $a^0a^0c^+$ tilting at room temperature. In addition to this tetragonal minority phase, a recent NMR experiment has provided direct evidence for the presence of an undistorted cubic phase.\cite{Groszewicz2014}\\

The domain structure has predominantly been investigated using transmission electron microscopy. Whereas an early study by \textcite{Soukhojak2000} reported the absence of domains in $\textrm{BaTiO}_3$-doped NBT, the presence of domains could later be verified. \textcite{Ma2010} have observed a strong composition dependence of the domain structure. For $x = 3.5\%$ and $x = 4\%$, rhombohedral domains of approximately 100~nm have been found to dominate. They are separated by domain walls, which are oriented parallel to $\{100\}$.\cite{Ma2013, Ma2010a} In contrast, significantly smaller domains with an average size of approximately 50~nm have been reported for $x = 4.5\%$.\cite{Yao2012} The tetragonal domains in NBT-3.6BT have not been observed directly. In the case of ceramics with $x = 6\%$, approximately 40\% of the grains possess a core-shell structure in which the shell region has $P4bm$ symmetry.\cite{Ma2010, Ma2010a, Ma2011} The shape of the tetragonal nanodomains in this region resembles thin platelets, which are oriented parallel to $\{100\}$. They may be embedded in an undistorted cubic matrix phase. In NBT, nanometer-sized tetragonal platelets\cite{Dorcet2008, Thomas2010, Matsuura2013, Ge2013a, Gorfman2015} (initially assumed to be monoclinic\cite{Kreisel2003, *Kreisel2004}) with orientation parallel to $\{100\}$ have been used as a model to explain the observed diffuse x-ray, neutron, and electron scattering features. The tetragonal platelets could be imaged directly by \textcite{Beanland2011a}, who have shown that, as opposed to $x = 6\%$ ceramics, the platelets are inhomogeneously distributed in the rhombohedral matrix that constitutes most of the sample volume. Consequently, these two possibilities for the arrangement and local environment of the tetragonal platelets have to be considered at intermediate compositions such as NBT-3.6BT. As a transmission electron microscopy experiment on $\textrm{K}_{1/2}\textrm{Na}_{1/2}\textrm{NbO}_3$-doped NBT-BT has clearly shown, such a two-phase structure can occur on the length scale of a few unit cells.\cite{Schmitt2010, *Schmitt2010a} The microstructure then approaches a modulated global structure such as the $a^-a^-a^-$/$a^-a^-c^+$ model proposed by \textcite{Levin2012} for NBT.\\

The various types of domains that were found in rhombohedral NBT are separated by different kinds of planar defects.\cite{Beanland2011} Ferroelectric nanodomains with a size of about 10 - 50~nm are separated by twin planes with preferential orientation parallel to $\{100\}$\cite{Dorcet2008, Levin2012, Ma2013} and local $a^-a^-c^+$ tilting.\cite{Dorcet2008a, Trolliard2008} The possibility of (multiple) rotational twinning within one rhombohedral nanodomain has been studied by Major\textit{ et al.}\cite{[] [ (submitted)] Major2016} They showed that a simple model featuring rotational twins as the only planar defects in a rhombohedral matrix is sufficient to reproduce the electric-field-dependent diffuse x-ray scattering features. In the case of NBT-3.6BT, the tetragonal and cubic phase fractions have to be taken into consideration, too. Since the $a^0a^0c^+$ tilts in adjacent layers of octahedra are decoupled along the polar $[001]$ axis, the formation energy for a stacking fault in the tilt sequence is very low.\cite{Meyer2015} Domains with different crystallographic symmetry can only be joined by incommensurate interfaces, which typically extend over two to three layers of octahedra.\cite{Groeting2012} Specifically, \textcite{Beanland2011} has suggested that the transition from $a^-a^-a^-$ to $a^0a^0c^+$ tilting occurs via an orthorhombic transition region with $a^-a^-c^+$ tilting. This possibility, as well as the previously mentioned modulated global structure model, will be discussed in Sec.\ \ref{sec:results}.\\

\begin{figure*}
\includegraphics[width=17.2cm]{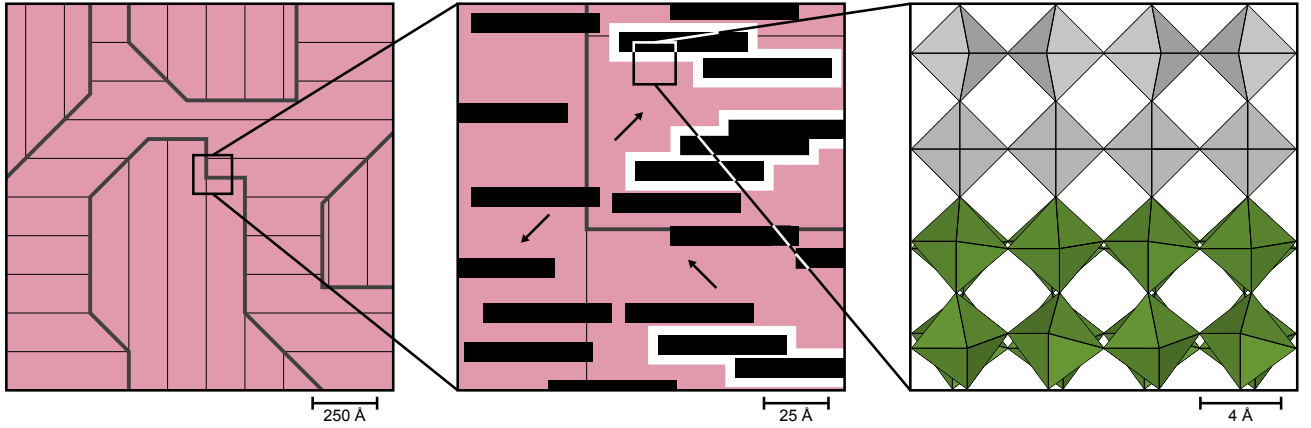}%
\caption{Two-dimensional model of the nanostructure of NBT-3.6BT at 310~K\@. Commensurate interfaces are drawn as thin lines, incommensurate interfaces as heavy lines. The first enlargement shows the distribution of tetragonal platelets (black) within the rhombohedral matrix phase [light red (grey)], separated by a cubic transition layer (white). A possible domain configuration in the rhombohedral region is also shown. The second enlargement shows the interface between the rhombohedral matrix and a tetragonal platelet on the atomic scale. See the text for further details.\label{fig:domain_structure}}
\end{figure*}

A two-dimensional model of the nanostructure of NBT-3.6BT is presented in Fig.\ \ref{fig:domain_structure}. In the first enlargement (center), the local distribution of the tetragonal phase is depicted in detail. Most importantly, it features nanometer-sized tetragonal platelets, some of which are separated from the rhombohedral matrix by a thin, tilt-free transition layer. The arrows indicate the local polarization direction in the rhombohedral domains. The second enlargement (right) shows a possible interface between the rhombohedral matrix and a tetragonal platelet. The tilt angle is exaggerated for clarity. Note the distortion of the oxygen octahedra in the lower two layers.\\

\begin{figure}
\includegraphics[width=8.6cm]{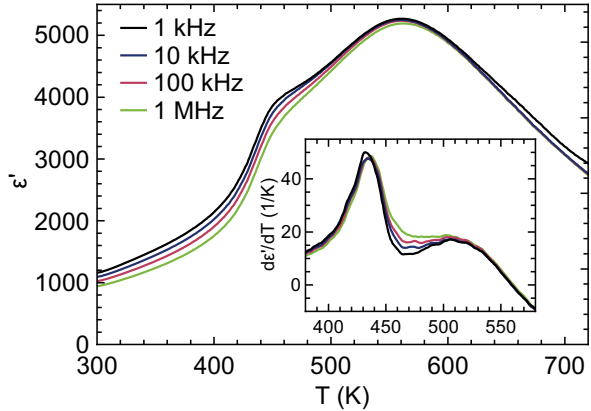}%
\caption{Temperature dependence of the dielectric function $\varepsilon'$ of unpoled NBT-3.6BT, measured at different frequencies between 1~kHz and 1~MHz. The frequency dispersion is clearly visible below approximately 500~K\@. The inset shows the derivative of $\varepsilon'$ with respect to temperature (smoothed).\label{fig:permittivity}}
\end{figure}

Figure \ref{fig:permittivity} shows the dielectric function of unpoled NBT-3.6BT as a function of temperature $T$ and measurement frequency. The data shown in this figure were measured on a single-crystal plate with orientation parallel to $\{100\}$. In the unpoled state, corresponding measurements on $\{110\}$- and $\{111\}$-oriented single crystals yield very similar results.\cite{Schneider2014} Like other relaxor ferroelectrics, NBT-3.6BT exhibits a clear frequency dispersion of the permittivity. In contrast to the prototypical relaxor $\textrm{PbMg}_{1/3}\textrm{Nb}_{2/3}\textrm{O}_3$, however, the temperature of the permittivity maximum $T_m$ does not depend on the measurement frequency in the case of NBT-3.6BT.\cite{Cross1987} Nevertheless, NBT-BT is commonly described as a relaxor system.\cite{Chen2008, Xu2008, Craciun2012} The depolarization temperature $T_d\approx 433~\textrm{K}$, which corresponds to the maximum of the derivative $d\varepsilon'/dT$ depicted in the inset, and $T_m\approx 563~\textrm{K}$ come rather close to the macroscopic phase transition temperatures reported by \textcite{Takenaka1991} $T_d$ and $T_m$ are, respectively, 30 and 20~K below the phase transitions. Whereas an indirect correlation clearly exists, it is likely not due to a direct causal link.\\

For lead-based relaxor ferroelectrics, a clear picture of the nanostructure has been established in order to explain their peculiar dielectric properties. It is based on the concept of polar nanoregions (PNRs).\cite{Bokov2006} In x-ray and neutron-scattering experiments, the PNRs have been associated with the diffuse scattering around fundamental reflections, i.e., reflections with integer $h$, $k$, and $l$.\cite{Gehring2012} These diffuse scattering features react to the application of an electric field to the sample, thus providing the link to its dielectric properties. It should be noted, however, that the association of this diffuse scattering with PNRs is not yet universally accepted.\cite{Bosak2012}\\

\begin{figure}
\includegraphics[width=7.6cm]{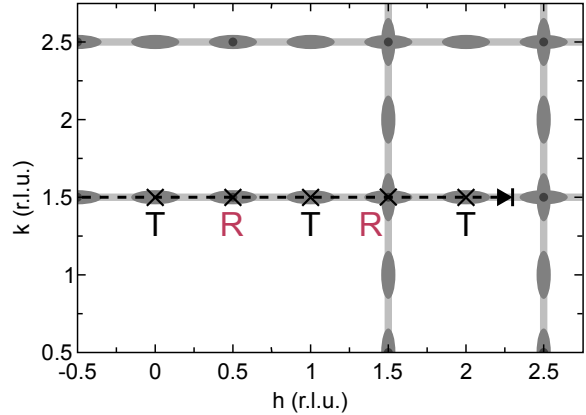}%
\caption{Schematic drawing of the scattering intensity distribution in the $l=\frac{1}{2}$ plane of reciprocal space at ambient temperature as known from diffuse x-ray scattering (selected features). $R$-type superlattice reflections (SRs) at $\frac{1}{2}\{ooo\}$ positions and $T$-type SRs at $\frac{1}{2}\{ooe\}$ stem from locally ordered tilt systems. Diffuse scattering streaks parallel to $h$ and $k$ are caused by planar stacking faults between nanoregions with uniform tilt. Neutron-scattering data were collected along $k=\frac{3}{2}$, $l=\frac{1}{2}$ as indicated by the dashed arrow.\label{fig:SR_diffuse}}
\end{figure}

In contrast, lead-free relaxor ferroelectrics such as NBT-BT behave very differently. As previous diffuse scattering experiments on NBT-BT with $x = 4\%$ have shown, it is the diffuse scattering in the scattering plane with $l=\frac{1}{2}~\textrm{r.l.u.}$ that exhibits an electric field dependence.\cite{Daniels2011, *Daniels2012} (Reciprocal space coordinates and scattering vectors are given in reciprocal lattice units based on the perovskite unit cell with $1~\textrm{r.l.u.} = 1.608~\textrm{\AA}^{-1}$.) In fact, all scattering planes with half-integer $l$ contain field-dependent diffuse scattering features. The most important scattering components in the $l=\frac{1}{2}~\textrm{r.l.u.}$ plane are schematically depicted in Fig.\ \ref{fig:SR_diffuse}. Apart from the previously mentioned SRs at $\frac{1}{2}\{ooo\}$ and comparatively diffuse SRs at $\frac{1}{2}\{ooe\}$, diffuse streaks connecting the $\frac{1}{2}\{ooo\}$ reflections parallel to $h$, $k$, and $l$ are the most prominent features.\cite{Balagurov2006, Dorcet2008, Daniels2011, *Daniels2012, Liu2012} The SRs, on the one hand, are due to the octahedral tilt within nanodomains with rhombohedral and tetragonal symmetry, respectively. On the other hand, the diffuse streaks are caused by planar defects in the nanostructure with orientation parallel to $\{100\}$ and an average distance of 20~nm.\cite{Major2016}\\

We characterize the temperature dependence of the disorder in NBT-3.6BT using elastic diffuse neutron scattering. The thermal evolution of the nanodomain structure is investigated with a special focus on correlation lengths. Finally, the observed features are related to the macroscopic permittivity.\\

\section{Experimental}

The data were collected at the thermal triple-axis spectrometer EIGER\cite{Wagner2006, *Blau2009} using a double focusing pyrolytic graphite (PG) monochromator and a horizontally focusing PG analyzer. The sample was an unpoled single-crystal boule with the composition $0.964\textrm{Na}_{1/2}\textrm{Bi}_{1/2}\textrm{TiO}_3-0.036\textrm{BaTiO}_3$. It was grown using the top-seeded solution growth method. The starting composition had previously been used to grow other crystals whose composition was verified using inductively coupled plasma atomic emission spectrometry. The weight of the sample was approximately 8.6~g and its shape resembled a disk with a diameter of approximately 17~mm and a thickness of approximately 8.5~mm. The symmetry axis had grown parallel to $[100]$. The sample was not polished.\\

The temperature was varied between 310 and 780~K using a small furnace. The sample was mounted in a small, thin-walled copper bucket to achieve high degrees of temperature stability and homogeneity. A thermocouple was placed very close to the sample, so that the sample temperature could be constantly monitored and stabilized.\\

Al$_2$O$_3$-based cement was used for fixing the sample on a copper wedge. This allowed us to conduct our measurements in the $\left(hk\frac{k}{3}\right)$ scattering plane as defined by the reciprocal lattice points $(100)$ and $(031)$. The Al$_2$O$_3$-based cement was also used for subsequently fixing the wedge inside the previously described bucket. The lid, made out of solid copper, closed the sample space and provided the mechanical and thermal connection between the bucket and the hot finger of the furnace. A second copper container was placed around the bucket as additional thermal shielding. It was surrounded by the vacuum container made of aluminium.\\

The temperature range from 310 to 780~K was covered in nine steps, focusing on the crystallographic phase transitions. At each step, the purely elastic scattering along $\mathbf{Q} = \left(h\frac{3}{2}\frac{1}{2}\right)$ with $-0.7 \leq h \leq 2.3$ was measured. Two further scans were performed without the sample at the highest and lowest temperature for later background subtraction.\\

The following data reduction steps were undertaken: A linear interpolation of the empty container scattering intensity was performed due to its weak temperature dependence. This background was subtracted from the measured data. If the measured intensity at any given data point was below three times the interpolated background intensity, this point was neglected in the further analysis as potentially unreliable. This mainly applied to $\left(h\frac{3}{2}\frac{1}{2}\right)$ points around $h = 1.0$ and $h = 1.5$, which coincide with two powder reflections of copper. The remaining reduced data were fitted with the sum of six Voigt peaks, representing the SRs, and a constant background. At $T \leq 480~\textrm{K}$, an additional Voigt component had to be introduced for a proper representation of the $R$-type SRs. In any given fit, the intensity of the SRs at $\mathbf{Q} = \bigl(\bar{\frac{1}{2}}\frac{3}{2}\frac{1}{2}\bigr)$ and $\left(\frac{1}{2}\frac{3}{2}\frac{1}{2}\right)$ was constrained to be equal, as were the Lorentzian widths for all $R$-type SRs and for all $T$-type SRs. The Gaussian width of all Voigt peaks was fixed at the calculated $\mathbf{Q}$ resolution of the spectrometer. This was estimated using the \textsc{iFit}\cite{Farhi2014, *Farhi2012}-based \textsc{ResLibCal} package. The Popovici algorithm\cite{Popovici1975} was used in the \textsc{ResLib} implementation. All least-squares refinements were carried out using \textsc{Fityk} 0.9.8.\cite{Wojdyr2010}\\

After the measurements along $\left(h\frac{3}{2}\frac{1}{2}\right)$, the sample was remounted on a different copper wedge. The new scattering plane $(hkk)$ was defined by the reciprocal-lattice points $(100)$ and $(011)$. Subsequently, additional purely elastic-scattering measurements were carried out along $\left(\frac{1}{2}kk\right)$ with $-0.25 \leq k \leq 1.25$ and along $\left(h\frac{1}{2}\frac{1}{2}\right)$ with $-0.25 \leq h \leq 1.75$. These measurements were made at 430, 480, and 570~K, respectively. The empty container scattering was measured at 480~K\@. The fits of the $\left(\frac{1}{2}kk\right)$ scans were made using one Lorentzian peak and a linear background. The peak position was held constant.\\

\section{Results and Discussion}\label{sec:results}

\begin{figure}
\includegraphics[width=8.6cm]{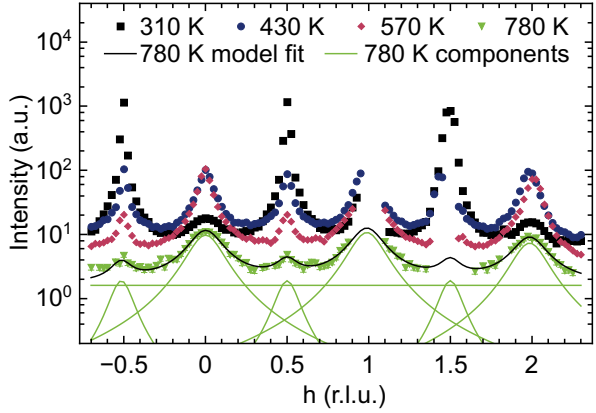}%
\caption{$\mathbf{Q}$ dependence of purely elastic neutron scattering along $\left(h\frac{3}{2}\frac{1}{2}\right)$ at selected temperatures after subtraction of interpolated empty container scattering. Error bars for the data recorded at 310, 430, and 570~K are smaller than the symbols and are not shown.\label{fig:H-scan}}
\end{figure}

The temperature and $\mathbf{Q}$-dependence of our elastic neutron scattering measurements are shown in Fig.\ \ref{fig:H-scan}. Missing data points around $h = 1.0$ and $h = 1.5$ were excluded from further treatment due to strong container scattering. At 310~K, the SRs at $R$-type $h = \pm \frac{1}{2}$ are strong and well-defined, whereas the scattering at $T$-type $h = 0$ and $h = 2$ is much weaker and broader. With increasing temperature, the intensity of the $R$-type SRs strongly decays, although they remain visible even at 780~K\@. The $T$-type SRs become stronger and sharper in the intermediate temperature range, but they are always visibly broader than their $R$-type counterparts. In addition, there is also a strong, $h$-independent component despite the previous background correction.\\

\begin{figure}
\includegraphics[width=8.6cm]{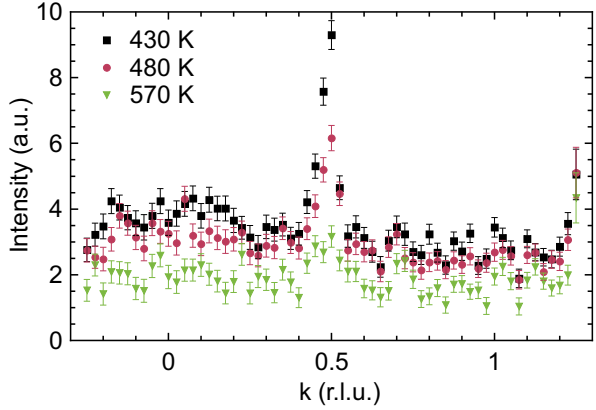}%
\caption{$\mathbf{Q}$ dependence of purely elastic neutron scattering along $\left(\frac{1}{2}kk\right)$ at 430, 480, and 570~K, after subtraction of the empty container scattering measured at 480~K\@.\label{fig:KL-scan}}
\end{figure}

In the measurements along $\left(\frac{1}{2}kk\right)$, which are shown in Fig.\ \ref{fig:KL-scan}, a temperature-dependent $M$-type superlattice reflection at $k = \frac{1}{2}$ is clearly seen. Its intensity shows the same temperature dependence as the $R$-type SRs, i.e., the intensity ratio is constant within the error bars at $\frac{I\left(\frac{1}{2}\frac{1}{2}\frac{1}{2}\right)}{I\left(\frac{1}{2}\frac{3}{2}\frac{1}{2}\right)} \approx 5-6\%$. This is a strong indication that both types of superlattice reflection originate from the same regions of the sample and that the $M$-type reflection is not due to higher order scattering from fundamental reflections such as $(111)$. However, we cannot exclude the occurrence of multiple scattering as discussed in Ref.\ \onlinecite{Gorfman2010}. Furthermore, the intensity ratio $\frac{I\left(\frac{1}{2}\frac{1}{2}\frac{1}{2}\right)}{I\left(\frac{1}{2}\frac{3}{2}\frac{1}{2}\right)}$ as predicted based on the model from Ref.\ \onlinecite{Usher2012} is much lower at about 0.2\% even for a bulk monoclinic sample. Therefore, we cannot draw any definitive conclusions regarding the crystallographic symmetry of the rhombohedral or monoclinic phase.\\

Second, we would like to emphasize the absence of the O-type $\left(\frac{1}{2}11\right)$ superlattice reflection in Fig.\ \ref{fig:KL-scan} at $k = 1$. \textcite{Levin2012} have argued that although $a^-a^-c^+$ tilting is present in NBT, the O-type SRs may not be detectable by electron scattering due to the low oxygen sensitivity of this probe. In the case of neutrons, the oxygen sensitivity is much higher, but still no $\left(\frac{1}{2}11\right)$ superlattice reflection is visible in our $\left(\frac{1}{2}kk\right)$ scans. Consequently, it appears unlikely that an orthorhombic phase with $a^-a^-c^+$ tilting is present in NBT-3.6BT in the intermediate temperature range. This supports our assumption that the symmetry transition between the coexisting $a^-a^-a^-$ and $a^0a^0c^+$ tilt systems does not occur via an intermediate phase with a$^-$a$^-$c$^+$ tilting. Instead, a completely tilt-free transition layer, as shown in Fig.\ \ref{fig:domain_structure}, seems to be the most probable possibility. Furthermore, our model provides a natural explanation for the detection of an undistorted cubic component by \textcite{Groszewicz2014}\\

Third, the background intensity in the purely elastic scans along $\left(\frac{1}{2}kk\right)$ with $0.8~\textrm{\AA}^{-1} \leq Q \leq 3.0~\textrm{\AA}^{-1}$ (Fig.\ \ref{fig:KL-scan}) is significantly weaker than the $h$-independent scattering component observed in the scans along $\left(h\frac{3}{2}\frac{1}{2}\right)$ (Fig.\ \ref{fig:H-scan}), where $2.5~\textrm{\AA}^{-1} \leq Q \leq 4.5~\textrm{\AA}^{-1}$. The background in the $\left(\frac{1}{2}kk\right)$ scans can thus be used to derive an upper bound for the weakly $\mathbf{Q}$-dependent scattering contributions that are present in the $\left(h\frac{3}{2}\frac{1}{2}\right)$ scans. The lower background intensity in the overlapping $Q$ range, and the reduction of the background with increasing $Q$ and $T$, which is evident in Fig.\ \ref{fig:KL-scan}, can also be observed in the scans along $\left(h\frac{1}{2}\frac{1}{2}\right)$ with $1.1~\textrm{\AA}^{-1} \leq Q \leq 3.1~\textrm{\AA}^{-1}$ (not shown). A tentative extrapolation to higher $Q$ using both $\left(\frac{1}{2}kk\right)$ and $\left(h\frac{1}{2}\frac{1}{2}\right)$ data correspondingly yields a weakly $\mathbf{Q}$-dependent background intensity that is significantly lower than the $h$-independent intensity found along $\left(h\frac{3}{2}\frac{1}{2}\right)$ in the measured temperature range. We thus assume that the $h$-independent scattering is dominated by the electric-field-dependent diffuse streak known from Ref.\ \onlinecite{Daniels2011, *Daniels2012}. Its intensity remains approximately constant up to 570~K and decreases upon further heating.\\

\begin{figure}
\includegraphics[width=8.6cm]{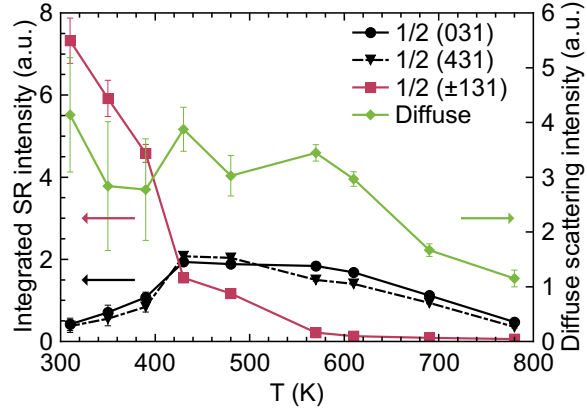}%
\caption{Temperature dependence of the integrated intensity of the SRs in Fig.\ \ref{fig:H-scan}, integrated over $\mathbf{Q}$ parallel to $h$. $T$-type SRs are drawn in black, $R$-type SRs in red (dark grey). Also shown is the diffuse scattering intensity that corresponds to the constant background in Fig.\ \ref{fig:H-scan}. Lines are a guide to the eye.\label{fig:elastic_intensity}}
\end{figure}

The detailed temperature dependence of both the superlattice reflection intensities and the elastic diffuse scattering intensity can be seen in Fig.\ \ref{fig:elastic_intensity}. The intensity of the SRs has been integrated parallel to $h$. The rhombohedral, tetragonal, and cubic phase fractions were estimated based on the integrated intensities of the SRs. The cubic phase fraction $f_C$ near room temperature was taken from \textcite{Groszewicz2014} as $f_C = (5.5 \pm 1.5)\%$. \textcite{Jones2002} have shown for NBT that the tilt angles in the rhombohedral and tetragonal phases are only slightly temperature-dependent over large temperature ranges. We thus assume that the temperature dependence of the structure factors $\lvert F\rvert$ can be neglected. Thermal displacements are not considered. For the $R$-type SRs, the structure factors were calculated using an $R3c$ model with a tilt angle of $6.52^\circ$ about the pseudocubic $[111]$ axis, which corresponds to the atomic positions given by Ranjan and Dviwedi.\cite{Ranjan2005} Since the tilt angle of pure NBT in the rhombohedral room temperature phase was given as $8.24^\circ$,\cite{Jones2002} the relative uncertainty of all tilt angles was assumed to be 20\%. A Bi$^{3+}$ position splitting of $0.30~\textrm{\AA}$ in the $\langle 01\bar{1}\rangle$ direction\cite{Thomas2005} was also taken into account. In the tetragonal phase, the Ti$^{4+}$ displacement along $[001]$ has no effect on the intensities of the SRs. The tilt angle has not been determined for NBT-3.6BT at elevated temperature, so that the model is based on the atomic positions provided by \textcite{Jones2002} for pure NBT at 673~K, which correspond to a tilt angle of $4.80^\circ$ about the pseudocubic $[001]$ axis.\\

\begin{figure}
\includegraphics[width=8.6cm]{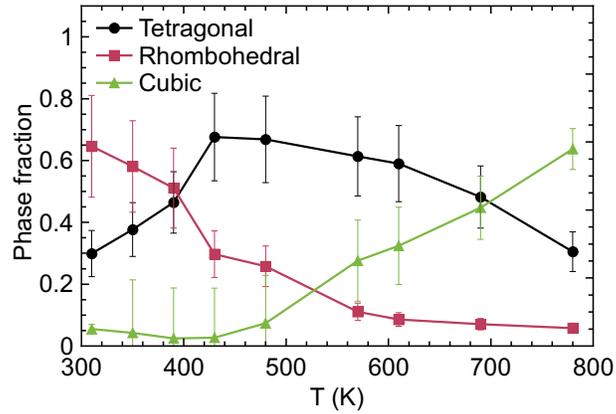}%
\caption{Temperature dependence of the phase fractions as estimated from the integrated intensity of the SRs. Lines are a guide to the eye.\label{fig:phase_fractions}}
\end{figure}

The measured intensities of the SRs were divided by the calculated $\lvert F\rvert^2$ and the Lorentz factors $L = \frac{1}{\sin 2\theta}$. A further correction was made to reflect the lower number of equivalent SRs in a monodomain single crystal compared to the assumed case of random orientation of the tilt domains. The resulting reduced intensity was assumed to be proportional to the respective phase fractions. It was calculated for both $T$-type SRs individually and subsequently averaged. To obtain the proportionality constant, the total phase fraction of the non-cubic phases had to be taken into account. With the assumption that this proportionality constant is not temperature-dependent, all phase fractions could be calculated as a function of temperature. The error bars in Fig.\ \ref{fig:phase_fractions} were determined by propagation of uncertainty.\\

Even though the measured temperature range includes the two macroscopic phase transitions from rhombohedral to tetragonal and from tetragonal to cubic, it is immediately apparent from Fig.\ \ref{fig:phase_fractions} that the rhombohedral and tetragonal phases are present throughout the investigated temperature range. The tetragonal phase fraction is very significant even at room temperature, which is over 100~K below the macroscopic rhombohedral to tetragonal phase transition. With increasing temperature, the rhombohedral phase fraction diminishes continuously. The cubic phase fraction appears to remain constant below about 430~K\@. Upon further heating, it grows linearly within the error bars. Nevertheless, at 780~K, which is about 200~K above the tetragonal to cubic phase transition, the actual cubic phase fraction is only around 64~\% due to the presence of both rhombohedral and tetragonal phases.\\

\begin{figure}
\includegraphics[width=8.6cm]{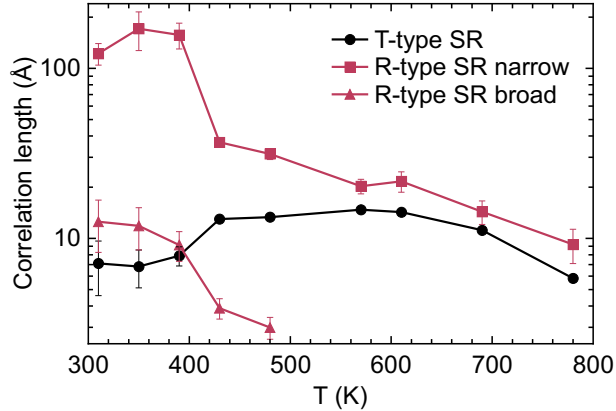}%
\caption{Temperature dependence of the tilt correlation length in the rhombohedral and tetragonal domains, as calculated from the profile parameters of the respective SRs. Two length scales of the rhombohedral domains have been obtained from the narrow and broad components of the $R$-type SRs. Lines are a guide to the eye.\label{fig:elastic_FWHM_corr_l}}
\end{figure}

The correlation lengths of the octahedral tilt order in the rhombohedral and tetragonal domains have been calculated from the Lorentzian broadening of the superlattice reflection profiles in Fig.\ \ref{fig:H-scan}. We assume that the correlations result from (possibly frozen) fluctuations, which lead to an Ornstein-Zernike form of the correlation function. The characteristic correlation length $\xi$ is then inversely proportional to the half width at half maximum $\Gamma$ of the corresponding scattering peak in reciprocal space. The mathematical relation is $\xi = \frac{a}{2\pi\Gamma}$, with $\Gamma$ given in reciprocal lattice units and $a$ representing the lattice parameter in real space. The resulting correlation lengths are shown in Fig.\ \ref{fig:elastic_FWHM_corr_l}. Since we used two Voigt components to fit the $R$-type SRs below 480~K, we obtained two characteristic length scales of the rhombohedral tilt order.\\

Figure \ref{fig:elastic_FWHM_corr_l} shows that the long-range correlations in the rhombohedral domains occur on a length scale of approximately 150~\AA\ up to 390~K\@. At higher temperature, a continuous reduction of the correlation length is observed. The length scale of the short-range correlations decreases from 13~\AA\ at 310~K to 3~\AA\ above 430~K, i.e., one pseudocubic unit cell. We assume that these short-range correlations are due to the overdamped $R_{25}$ phonon, as proposed by Ge\textit{ et al.}\cite{Ge2013}\\

On the other hand, the correlation length of the tetragonal domains increases by a factor of two around the rhombohedral to tetragonal phase transition and continues to increase to its maximum of roughly 15~\AA\ around 600~K\@. It is particularly striking that the correlation length of the tilts in the tetragonal domains remains below 15~\AA\ even when the sample is macroscopically tetragonal. This can be understood if the tetragonal domains are envisaged as small platelets in a rhombohedral or cubic matrix.\\

\begin{figure}
\includegraphics[width=8.6cm]{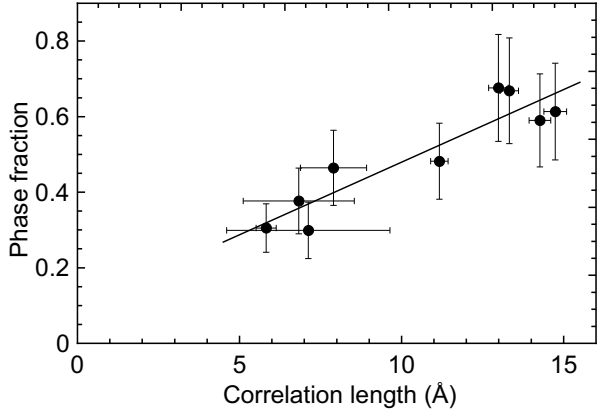}%
\caption{Scatter plot of the tetragonal phase fractions and tilt correlation lengths at all temperature points. The solid line is a linear fit to the data.\label{fig:corr_T_vs_corr_l}}
\end{figure}

The phase fraction and tilt correlation length seem to be qualitatively correlated in both phases (compare Figs.\ \ref{fig:phase_fractions} and \ref{fig:elastic_FWHM_corr_l}). It turns out that the correlation coefficients $r_{xy}$ are 0.92 for the rhombohedral phase (narrow component), 0.996 for the rhombohedral phase (broad component), and 0.92 for the tetragonal phase. Eliminating the influence of temperature yields partial correlation coefficients $r_{xy\setminus T} = 0.80$ and 0.95, respectively, for the rhombohedral phase. Both correlation coefficients decrease slightly when the effect of temperature is eliminated, which shows that a causal relation between the correlation length and the phase fraction may be present. On the other hand, $r_{xy\setminus T} = 0.93$ for the tetragonal phase, which is higher than $r_{xy}$ and thus a strong indication that the phase fraction indeed depends on the correlation length. This dependence is shown in Fig.\ \ref{fig:corr_T_vs_corr_l}.\\

To derive a possible explanation for this linear dependence, let us briefly reconsider the dependence of the octahedral tilting on the chemical short-range order. The influence of chemical short-range order on the stability of different tilt systems has been shown for pure NBT by density-functional-theory calculations.\cite{Groeting2012} A crucial result of \textcite{Groeting2012} was that at ambient pressure, the local cation configurations designated as ``110'', ``11-01xy'', and ``111'' favor out of phase $a^-a^-a^-$ tilts, whereas $a^-a^-c^+$ tilts are favored by local ``10-01z'', ``all3+1'', and ``001'' cation ordering. In agreement with Levin and Reaney,\cite{Levin2012} \textcite{Groeting2012} propose the existence of chemically ordered nanoregions with $a^-a^-c^+$ tilting, embedded in a rhombohedral matrix. As we have shown, no O-type SRs are present in the case of NBT-3.6BT (see Fig.\ \ref{fig:KL-scan}). We therefore assume that the in-phase tilting in NBT-3.6BT occurs in tetragonal nanoregions with $a^0a^0c^+$ tilt order. This difference between NBT and NBT-3.6BT seems plausible considering that NBT-3.6BT is closer to the tetragonal phase field and, according to Ma\textit{ et al.},\cite{Ma2013} even in a different phase field than pure NBT.\\

Since the local cation ordering is stable as long as cation diffusion is not activated, the tetragonal nanoregions remain fixed in space and it can be assumed that their number density does not change significantly with temperature. It can thus be said that the tetragonal platelets seem to be chemically pinned. In addition to this effect of the local Na/Bi ordering, the tilt structure depends strongly on the $\mathrm{Ba}^{2+}$ concentration,\cite{Groeting2014} so that chemical inhomogeneity could add to the pinning effect in NBT-BT. Coming back to the interpretation of Fig.\ \ref{fig:corr_T_vs_corr_l}, it seems likely that the thickness of the tetragonal platelets varies most strongly with temperature, whereas their other dimensions remain constant. Thus, the volume of the platelets would be a linear function of their thickness. Neglecting the possibility of stacking faults along the tetragonal $[001]$ direction, the thickness of the platelets corresponds to the tilt correlation length. If the number density of the tetragonal platelets is finally assumed to be constant, a linear dependence of the phase fraction on the tilt correlation length would result.\\

On this basis, a clear picture of the nanostructure can be drawn. The tetragonal platelets are distributed inhomogeneously in the crystal, as has been shown for pure NBT.\cite{Dorcet2008, Beanland2011a} At low temperatures, they are surrounded by large rhombohedral domains. The incompatibility of the tilt systems leads to a continuous transition, which in some cases involves a thin cubic layer at the interface. Due to the reduced stability of the rhombohedral tilt system at intermediate temperatures, the rhombohedral domains shrink, while the tetragonal platelets grow. At high temperatures, some rhombohedral domains disappear and the tetragonal platelets begin to shrink. The cubic interface layers, which now surround all platelets, grow to coalesce into a continuous matrix.\\

\begin{figure*}
\includegraphics[width=17.2cm]{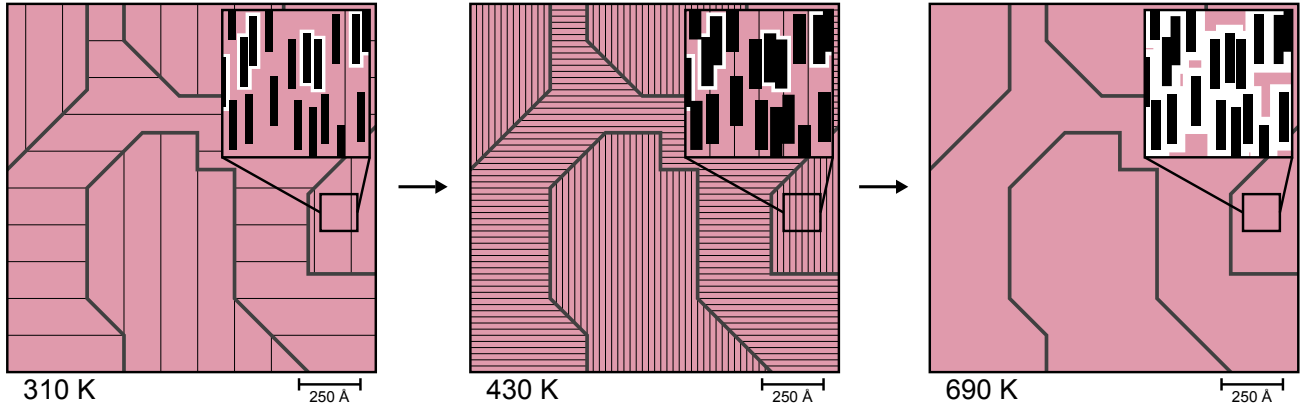}%
\caption{Proposed model for the thermal evolution of the nanostructure sketched in Fig.\ \ref{fig:domain_structure}. The temperature-dependent shape of the tetragonal platelets and the tilt-free intermediate layers is shown in the insets. The graphical representation is the same as in Fig.\ \ref{fig:domain_structure}. See the text for further details.\label{fig:T-dep_domains}}
\end{figure*}

Figure \ref{fig:T-dep_domains} shows a model for the thermal evolution of the domain structure that was sketched in Fig.\ \ref{fig:domain_structure}. The low cubic phase fraction at low to medium temperatures indicates that the tilt-free layers around the tetragonal platelets are very narrow, i.e., about one unit cell thick. In many cases, the tilt transition occurs on such a small length scale that all layers are distorted and a tilt-free layer does not occur at all. At high temperature, the cubic phase becomes dominant and the rhombohedral phase disappears. This means that the observed reduction of the correlation length in the rhombohedral phase upon heating is only partly due to domain fragmentation within the rhombohedral regions, which has mainly been observed at low to intermediate temperatures.\cite{Dorcet2008a} The continuous growth and coalescence of the tilt-free regions clearly leads to an additional reduction of the rhombohedral domain size due to the inward movement of the phase boundaries.\\

\begin{figure}
\includegraphics[width=8.6cm]{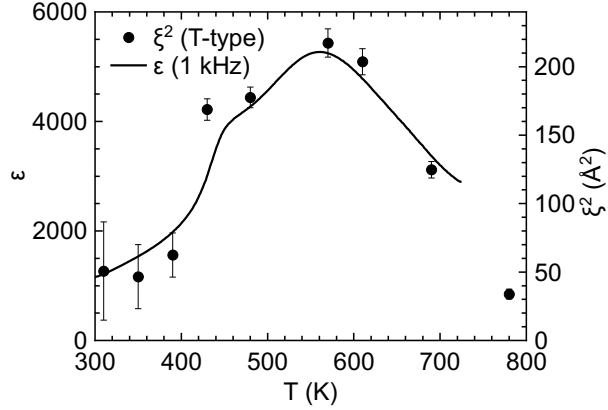}%
\caption{Temperature dependence of the dielectric permittivity $\varepsilon$ (from Fig.\ \ref{fig:permittivity}) and the squared correlation length $\xi^2$ of the tetragonal domains (from Fig.\ \ref{fig:elastic_FWHM_corr_l}).\label{fig:corr_RT}}
\end{figure}

A very interesting structure-property relationship can be observed in the correlation between the dielectric permittivity $\varepsilon$ and the squared correlation length of the tetragonal domains. This is shown in Fig.\ \ref{fig:corr_RT}.\\

It is clearly seen in Fig.\ \ref{fig:corr_RT} that the temperature dependences of the permittivity and the squared correlation length are very similar. This shows that the thickness of the tetragonal platelets may be of crucial importance for the dielectric properties of NBT-based relaxors. If the polarizability were constant throughout the tetragonal platelets, a linear dependence of $\varepsilon$ on the volume fraction and thus on $\xi$ (see Fig.\ \ref{fig:corr_T_vs_corr_l}) would be expected. The observation that $\varepsilon$ is in fact proportional to $\xi^2$ could indicate that the local polarizability increases linearly with increasing distance from the undistorted cubic phase, i.e., with increasing octahedral tilting. Then $\varepsilon$, which is proportional to the volume integral of the local polarizability, would be proportional to $\xi^2$, as we observe. This is consistent with the observation that the tetragonal phase fraction increases upon poling and exhibits a more pronounced field-induced distortion.\cite{Jo2011a, Ge2013a, Hinterstein2015} A similar, qualitative correlation between the composition-dependent volume fraction of the tetragonal phase and the piezoelectric properties of NBT-BT has previously been proposed in Ref.\ \onlinecite{Yao2012}. Since we could show that the volume fraction of the tetragonal phase is proportional to its correlation length, the findings of \textcite{Yao2012} are also consistent with the correlation shown in Fig.\ \ref{fig:corr_RT}.\\

\section{Conclusions}

We have characterized the temperature dependence of the octahedral tilt order in NBT-3.6BT using elastic diffuse neutron scattering. In order to determine the volume fractions of the different octahedral tilt systems, we have investigated the intensity of the associated superlattice reflections. We have found no indication that an orthorhombic component is present. The phase fractions of the rhombohedral, tetragonal, and cubic components depend strongly on the temperature: Close to room temperature, the rhombohedral phase dominates, but the phase fraction decreases with increasing temperature. The tetragonal phase dominates in the intermediate temperature range. Since the reduction of the tetragonal phase fraction at high temperature is not very pronounced, the macroscopically cubic phase still contains over 30~\% of the tetragonal phase, 200~K above the $P4bm \rightarrow Pm\bar{3}m$ phase transition.\\

Furthermore, we have derived the correlation lengths of the rhombohedral and tetragonal domains from the superlattice reflection profiles. We have found it necessary to include two components of the rhombohedral SRs in the fitting model below 480~K, leading to two correlation lengths on different length scales. Both correlation lengths of the rhombohedral domains decrease continuously with increasing temperature. The tetragonal domains undergo only small size changes between 6~\AA\ and 15~\AA, and thus remain consistently smaller than the rhombohedral domains. Notably, the correlation length of the tetragonal domains exhibits the same temperature dependence as the tetragonal phase fraction. These results lead us to the conclusion that the nanostructure at ambient temperature features chemically pinned tetragonal platelets embedded in the rhombohedral matrix, often separated by a cubic intermediate phase. The thickness of the tetragonal platelets changes with temperature. The thickness of the cubic intermediate layers increases above the $R3c \rightarrow P4bm$ phase transition and leads to the coalescence of the cubic regions at high temperature, when the cubic phase is the dominant component of the matrix. The additional short range order in the rhombohedral phase below the macroscopic tetragonal to rhombohedral phase transition may be due to the soft $R_{25}$ phonon mode.\\

We have also found clear indications that the dielectric permittivity is a function of the tetragonal platelet thickness squared. This correlation leads us to the conclusion that the tetragonal phase becomes more polarizable with increasing distance from the tilt-free interfaces. However, the composition dependence of the dielectric properties shows that a bulk tetragonal phase exhibits a weaker response than a two-phase mixture.\cite{Jo2011a, Garg2013} Thus, the lattice strain induced by the surrounding rhombohedral matrix may facilitate the field-induced distortions whithin the tetragonal phase, so that a two-phase system with a high density of platelets with intermediate thickness could exhibit the best dielectric properties in the NBT-BT system. Due to the chemical pinning effect, the size and distribution of the tetragonal platelets can probably be influenced by the local chemical order. This means that controlling the chemical order may turn out to be an additional way of tailoring the dielectric properties of NBT-BT.\\

As a final remark, we would like to point out that our studies of the static structural features provide indications towards the atomistic origin of the temperature dependence of the dielectric permittivity. However, the frequency dependence of the permittivity at lower temperatures remains unexplained. Inelastic-scattering studies might help to shed some light on the underlying dynamic processes.\\

\begin{acknowledgments}
The authors would like to acknowledge Daniel Rytz for the synthesis of the investigated single crystal. The authors are also indebted to Deborah Schneider for providing the permittivity data and insightful discussions. This work is based on experiments performed at the Swiss spallation neutron source SINQ, Paul Scherrer Institute, Villigen, Switzerland. This research project has been supported by the European Commission under the 7th Framework Programme through the ``Research Infrastructures'' action of the ``Capacities'' Programme, NMI3-II Grant number 283883. This work was funded by the Deutsche Forschungsgemeinschaft (DFG) under SFB 595 ``Electrical Fatigue in Functional Materials.''
\end{acknowledgments}

\end{document}